Marianne Gunderson,
PhD candidate in Digital Dulture
Department of Linguistic, Literary and Aesthetic Studies, University of Bergen
marianne.gunderson@uib.no / marianne.gunderson@gmail.com

# Visions of augmented reality in popular culture:
*Power and (un)readable identities when the world becomes a screen*

*(Translation by Semantix, originally published in Norwegian)*

**Abstract**

Augmented reality, where digital objects are overlaid and combined with the ordinary visual surface, is a technology under rapid development, which has long been a part of visions of the digital future. In this article, I examine how gaze and power are coded into three pop-cultural visions of augmented reality. By analysing representations of augmented reality in science fiction through the lens of feminist theory on performativity and intelligibility, visibility and race, gendered gaze, and algorithmic normativity, this paper provides a critical understanding of augmented reality as a visual technology and how it might change or reinforce possible norms and power relations. In these futures where the screen no longer has any boundaries, both cooperative and reluctant bodies are inscribed with gendered and racialised digital markers. Reading visions of augmented reality through feminist theory, I argue that augmented reality technologies enter into assemblages of people, discourses, and technologies, where none of the actors necessarily has an overview. In these assemblages, augmented reality takes on a performative and norm-bearing role by forming a grid of intelligibility that codifies identities, structures hierarchical relationships, and scripts social interactions.

**Introduction**

We are surrounded by digital information. The internet is often the first place we go to contact friends, look for love, be entertained, check facts, find out what is happening in the world or our local community, and participate in political discussions. More and more areas of our lives are being incorporated into digital networks and integrated with algorithmic tools, which are supposed to help us keep track of what we do and choose what suits us best, be it exercise, eating habits, shopping or entertainment. Similar developments are also occurring within essential societal functions such as policing, health services and education. These technological interventions are not neutral: they are both shaped by and shape power relations and structural inequalities in the communities and lives they encounter. Aspects such as gender, skin colour, and class are coded into these structures with varying degrees of subtlety, which influences how different people will experience or be treated by different digital systems. For now, we primarily use screens to interact with these systems, but what if this was no longer necessary? What if digital information was available directly in our field of vision? This is the vision that augmented reality promises – or threatens – to fulfil.

Augmented reality (often abbreviated as AR) merges virtual objects with a physical environment, where digital objects are superimposed on to or combined with physical objects in the ordinary visual surface (Azuma 1997). This technology has long been part of visions of our digital future but has yet to become part of everyday life for most people. As I will discuss in this article, there is much to suggest that augmented reality will become a much more integral phenomenon in our lives in the future. Major global technology companies such as Apple, Google, Microsoft and Facebook have all invested large sums in developing augmented reality solutions for ordinary consumers, and several products are expected to be launched in the coming years. It is therefore relevant to investigate how this technology might influence, and be influenced by, existing power relations, norms and identities. How is augmented reality gendered? What kind of relationships and subjectivities will be encoded into the visual experience when the world becomes a screen? How will we see gendered and racialized bodies in augmented reality? These are questions that must be addressed if we want to have an established framework of understanding and critical awareness of this technology before it becomes part of our everyday lives.

Feminist theory has a rich history of producing critical thinking about the correlations between visibility, gaze and power in fields such as media research, theory of science and digital culture. While much has been written about virtual reality from a feminist perspective and field of research long before this technology was widely available to the general public (Hawthorne 1999; Hayles 1999; Sophia 1992), there is currently little corresponding research on augmented reality, despite the fact that this technology also has a long history. There is therefore a need for research that takes a critical look at these aspects of augmented reality, and this article aims to help bridge this knowledge gap and form a basis for further discussion about the power dimensions in augmented reality.

Although there is currently little research on this technology's social or societal consequences, science fiction has widely explored these aspects in one popular cultural genre. In this article, I will analyze three different visual science fiction works: the film *Anon* (Niccols 2018), the TV series *Brave New World* (Morrison, Taylor, and Wiener 2020), and the short film *Sight* (May-raz and Lazo 2012). Each of these addresses how augmented reality produces or reproduces identity categories, power relations, and normativity.

Much of the academic work on science fiction in feminist theory has focused on works either made by women, where gender is clearly a topic, or that are otherwise clearly feminist (Haran and King 2013). The works discussed in this article have not been included because they are distinctly feminist contributions to the genre, although they are readily available for feminist reading. Instead, they have been chosen partly because they are so unnoteworthy: They are relatively mainstream entertainment science fiction, created to compete for viewers on streaming platforms such as Netflix, HBO and YouTube. Their lack of originality is precisely what makes these works interesting, to the extent that they can problematize the immediate, general notions of how augmented reality might manifest itself in the future. The key research question that has directed the present article is: What notions of gaze and power are encoded into future visions of augmented reality? My aim in reading these works is therefore to investigate how, in these visions of the future augmented reality helps maintain or erode power structures, how it affects identity categories, and whom it empowers and whom it leaves powerless – and what these imaginings can teach us about the challenges augmented reality may entail in reality.

In my analysis of these works, I show how science fiction problematizes this technology by showing us how normative identities, existing hierarchies and social divisions can be coded into people's field of vision using augmented reality, and how this might serve to make these structures more rigid and pervasive. This article argues that augmented reality can mobilize existing norms and structures, cement power relations, and constrict possible positions of subjectivity. By constructing a dialogue between science fiction and feminist theory on performativity and intelligibility, visibility and race, gendered gaze and algorithmic normativity, this paper provides a critical interpretation of augmented reality as a visual technology, and takes a critical look at how this technology might change or reinforce existing norms and power relations.

**From fiction to reality**

A first question we must ask is how can we understand the social consequences of a technology that has yet to be realised? Looking at how augmented reality works today in real life will provide a very meagre basis for analyzing how this technology might behave in the future, as long as it still has relatively limited areas of application and distribution. The fact that the technology is still relatively young means that there is also limited academic work directly addressing how augmented reality might affect social interactions and power relations. Imagining the possible consequences of an innovation will always involve an element of speculation, and in this area no genres have been as productive as science fiction. In the essay *The SF of Theory*, Istvan Csicsery-Ronay (1991) points out that the visions of the future in science fiction span two voids: they bridge the gap between what is technically possible in the present and what is imagined that will be achievable in the future, and the ethical gap between what might happen, realistically, and what ought to happen, morally. This has made science fiction a rich source of social criticism, and feminist theory has a solid tradition of thinking with, and through, science fiction (Haran and King 2013; Haraway 1989). A speculative approach often entails a component of alienation or defamiliarization of social relations, and can pave the way for notions of new normativities, making the texts highly relevant to feminist thinking (Melzer 2006). For the feminist cultural theorist Teresa de Lauretis (1980), science fiction as a genre is outstanding in its ability to present a cultural reality shaped by technology at all levels – both economically and psychologically, publicly and privately, while Donna Haraway (2000) states that for her, science fiction is political theory. Ingvil Hellstrand (2016, 2017) points out that science fiction can show us possible (and impossible) futures, where the boundaries of familiar cultural representations or "imaginaries" about what it means to be human can be renegotiated. Together, this helps turn the spotlight on things that are usually taken for granted. For the purpose of my analysis, it was therefore natural to turn to science fiction as source material to reflect on how augmented reality vision might be imagined to influence the boundaries that define possible categories of identity and subjectivity.

As a genre, science fiction has a growing number of works that portray variants of this technology, but often without augmented reality vision being directly thematicized in the narrative. In the works I have chosen to analyse in this article – the film *Anon* (2018), the TV series *Brave New World* (2020) and the short film *Sight* (2012) – augmented reality plays a central role both in the aesthetic viewing experience and in the story that unfolds. By including works from different visual media or genre formats, this selection helps illustrate some of the breadth of representations of this technology in different genres. At the same time, these works are quite diverse in terms of both theme and genre: *Brave New World* is a re-adaptation of Aldous Huxley's classic, dystopian novel (2006 [1932]); *Sight* is a short-storyesque tale of a date, and *Anon* adheres to the film noir genre. The fact that the three works all have a different thematic focus and format means they also explore different situations and aspects of augmented reality.

At the same time, they are each dystopian in their own way, in the sense that they present a pessimistic vision of the future, where augmented reality is a means to oppress or use force against groups or individuals (Moylan 2013). Dystopian science fiction has a long tradition of exploring problematic aspects of new technologies, especially when it comes to surveillance technology and machine vision (Marks 2015). This makes science fiction an excellent laboratory for the development of technology criticism, because it nurtures thought experiments that explore what can go wrong in the interface between augmented reality and existing forms of inequality, normativity, and oppression. In the absence of empirical data on what will happen when this technology becomes universal, these narratives can play out scenarios that clarify what might be at stake with the introduction of new technologies and how these technologies might have unforeseen social consequences at the individual and group level.

**Augmented reality**

The idea of a technology that can integrate digital information into the field of vision is not as recent as one may think. The first prototype of an augmented reality device was designed at Harvard in 1968. Ivan Sutherland and Bob Sproull's stationary prototype could create three-dimensional graphics in 40 per cent of the user's field of vision, where both the real world and the synthetic information were visible at the same time. The projections could either float in thin air or be aligned with maps, desks or the keys on a typewriter (Billinghurst 2015). In the 1980s and 1990s, the concept was further developed to assist pilots in poor weather conditions and for use in surgery and industrial construction. Towards the end of the 1990s and early 2000s, several helmets with augmented reality glasses were developed, which could be combined with GPS systems or computer games to give the user a blended reality experience. However, these systems were hampered by their cumbersome design, weight and battery capacity (Billinghurst 2015).

The mobile phone game Pokémon Go introduced augmented reality as part of everyday gaming for many people, but here the digital information is displayed on the phone screen when the user points the phone camera towards a Pokémon. The best example of augmented reality integrated into the visual field in recent times is smartglasses that can display digital information in the wearer's field of vision. Although these devices are now in use in a number of specialised contexts, none has yet been widely embraced as consumer technology in the same way as mobile phones or smartwatches. Nevertheless, there is much evidence that the major technology companies have great expectations for this type of technology. A new model of Google Glass, Google's smartglasses, which were launched in 2013, is now available with AR functionality, and Microsoft has released a new version of its HoloLens AR glasses system. These glasses are currently mainly marketed as tools for companies in industry, manufacturing and logistics, but consumer versions are also underway. Facebook has announced its vision to create AR glasses that "will fundamentally change the way we interact with and experience the digital world" (Facebook 2020), with the first model due to launch in 2021 (Horwitz 2020). The Chinese mobile technology company Oppo has announced similar plans (Mehta 2020). Apple is also reported to be in the process of developing an AR eyewear concept (Rubin 2019). Time will tell whether these projects will succeed and how they will interact with social media, artificial intelligence and other digital structures. However, there is no doubt that significant resources are being invested in making augmented reality a part of our everyday lives in the future.

**Readable and unreadable identities**

The film *Anon*, produced for Netflix in 2018, takes place in a near future where everyone has optic nerve implants that record everything they see and do. At the same time, the implants give people access to an augmented reality where their field of vision is filled with information about the object or person they are looking at. In the opening scene, we follow Sal, a police officer, as he walks down a street in an unnamed city. The scene is colourless and dark, while everything he looks at is surrounded by white lines and text. All the passers-by have a white box around their head showing their name, age and profession; cars are accompanied by text boxes showing the car brand, model and what music the driver is listening to; even dogs are tagged with information about their name and race. He notices a young woman, with the label UNKNOWN - ERROR above the square around her head. This unidentified woman later becomes the prime suspect in an investigation into a series of murders committed by an unseen perpetrator.

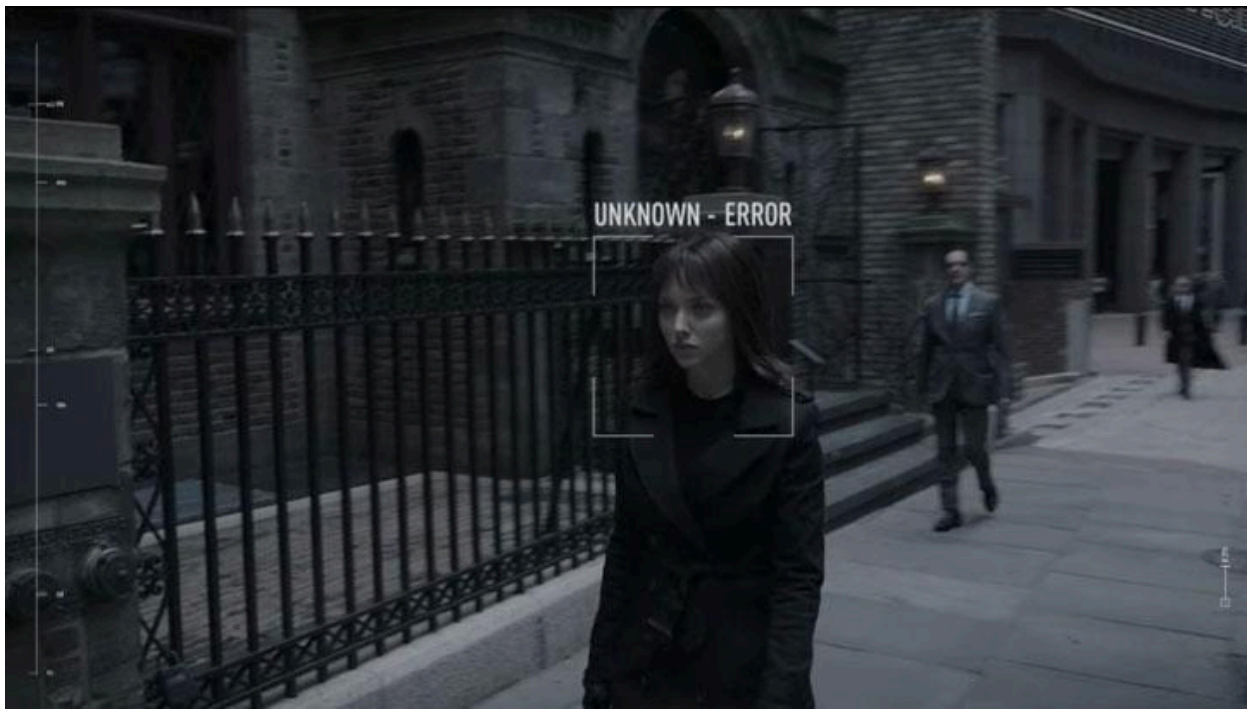

Figure 1, screenshot from Anon

The augmented reality in *Anon* means that no one can hide. You are no longer an anonymous face in the crowd; your name, age and profession are on display wherever you go. In 1995, Sherry Turkle argued that virtual identities gave people the opportunity to disguise or conceal aspects of their identity that caused them to be discriminated against in the rest of their lives, enabling experimentation with identities. Since then, more and more social media, with Facebook at the helm, have started requiring users to use their real names when creating an account. This spelled the end of the utopian visions of an online community where people could escape the identities associated with their physical bodies (Nakamura 2015). The digital network in *Anon* can be seen as a logical endpoint for this development. People's physical bodies are constantly framed and contextualised by digital information, which is always on display wherever you go. This information is not a creative self-presentation where people can choose which identity categories they wish to display; static tags show their public identity: name, age, and occupation. The problem arises when Sal crosses path with a person who is unreadable, a woman who does not reveal who she is at first glance.

From this perspective, *Anon* represents the idea that certain aspects of people's identities ought to be visible to anyone who glances at their face. The French philosopher Elsa Dorlin (2016) points out that visibility has become essential to how Western states operate. With an analysis of the discourse around the niqab ban in France, she shows how state power works through an obligation to present oneself, to be visible and transparent. At the same time, invisibility becomes a deviant behaviour that is excluded and punished. Dorlin points out how the recognizability of the human face has become the source of individuals' identities in their encounters with the state. The first people to undergo this form of control were criminals, who were photographed on arrest, and this has since become the norm for the rest of the population as well (Dorlin 2016:250). However, while Dorlin talks about the naked face as a carrier of identity, in *Anon,* this has acquired a mechanical element. The digital frame around the people's heads demonstrates that identity still follows the face, but verification is now done by a program that immediately interprets and confirms everyone's identity.

When machine learning attempts to describe the characteristics of an image (for example, to determine whether a portrait is of a man or a woman), the entire image is reduced to the markers and labels it has been programmed to use, leaving little room for ambivalence or uncertainty (Kronman 2020). In *Anon*, where these labels become visible, obligatory markers of identity, they function like what Judith Butler calls a "grid of intelligibility" (Butler 2004:45) – a norm system that defines the framework for what is displayed and what is not displayed in the social sphere (ibid.:52). In this context, augmented reality serves as a kind of "technological filter" (Rettberg 2016) that reduces faces to what is relevant for the control function by highlighting the public individual identity, while other identity markers remain unmarked in the digital layer. In the same way as gender norms, according to Butler, establish which subjects are culturally intelligible and thus recognizable, in *Anon* it is the digital filter that determines who is intelligible and recognizable.

The layer of digital information that supplements the field of vision entails a constant connection to a digital world where everything you do leaves permanent, traceable marks. The labels that make citizens intelligible to one another also make them intelligible to the state. When they are part of the digital system that sustains the augmented reality, the citizens become part of what Jacques Rancière (2004) describes as the "distribution of the sensible", that is, the complex of individuals in a society that can be sensed and are therefore definable bodies. Making identities instantly sensible also makes them susceptible to control and use of force, and visibility becomes a political resource. In *Anon*, this is further elaborated upon by the augmented reality being part of a centralized system that monitors the citizens, thereby making them predictable and compliant citizens. Lisa Nakamura (2015:221) points out that surveillance is not only an act of observing but that it also transforms the body into a social actor, where some bodies are classified as normative and legal, and others are classified as illegal or undesirable. By evading identification via the augmented reality system, the woman, who in *Anon* remains nameless, does what Rancière (2013) claims that those in power fear: She produces disorder in the established classification system and evades society's control. Her ability to remain unclassified makes her an unreadable figure, and thereby a destabilising threat to the established order in society.

**Biometric hierarchy**

The TV series *Brave New World* (2020) takes this concept one step further. Based on the novel by Aldous Huxley (2010) of the same title, the series depicts a society in which the citizens are biotechnically designed for the roles they are meant to fulfil in society, and where dissatisfaction and other unpleasant feelings are kept under control through the use of mood-altering drugs. Here, too, all the citizens are accompanied by a visual digital inscription that denotes a stable, static and pre-defined identity, but in this case the labels, aptly named "signifiers", are explicitly linked to the individual's position in a biologically determined hierarchy. You can tell at a glance whether the person you are looking at is an Alpha, Beta, Gamma, Delta or Epsilon. These labels describe their biological genetic material and corresponding abilities and weaknesses, defining their position in the genetic caste system. While the categories were present in the original novel, the augmented reality digital labels were introduced in this adaptation, and thus reflect the present era's vision of our digital future. At the same time, all the citizens are connected to a central network controlled by an artificial intelligence known as Indra. One striking feature in this adaptation is that they have chosen to display the individuals' position and status through a network-based augmented reality, rather than other characteristics. What difference does it make when these categories are displayed as digital information overlaid in the characters' field of vision?

At first glance, the augmented reality in *Brave New World* seems to serve a similar function as in *Anon*. In both films, the social categories are rendered visible through augmented reality. However, in *Brave New World,* the focus is not on individual surveillance and control; instead the genetic "truth" about the individual's body is inscribed in the digital layer, where the castes act as a biometric classification system. Simone Browne has argued that biometrics, defined as "a means of body measurement that is put to use to allow the body, or parts and pieces and performances of the human body, to function as identification" (2015:102), has a long history that is closely intertwined with eugenic principles and so-called racial hygiene practices. In *Dark Matters: On the Surveillance of Blackness* (2015), Browne traces biometrics back to the practice of branding black people in the transatlantic slave trade. She points out that branding contributed to black bodies being classified as tradable property while limiting and defining opportunities for black people's lives. According to Browne (2015), today biometrics still helps codify racialized information inscribed digitally onto specific bodies, for example by black bodies being marked as deviant or criminals in identification systems based on facial recognition, body scans or DNA. In particular, she points to the dangers associated with what she calls "noncooperative biometric tagging", in which biometric monitoring technologies help produce truths about racialised bodies (2015:128) using augmented reality devices, such as Google Glass.

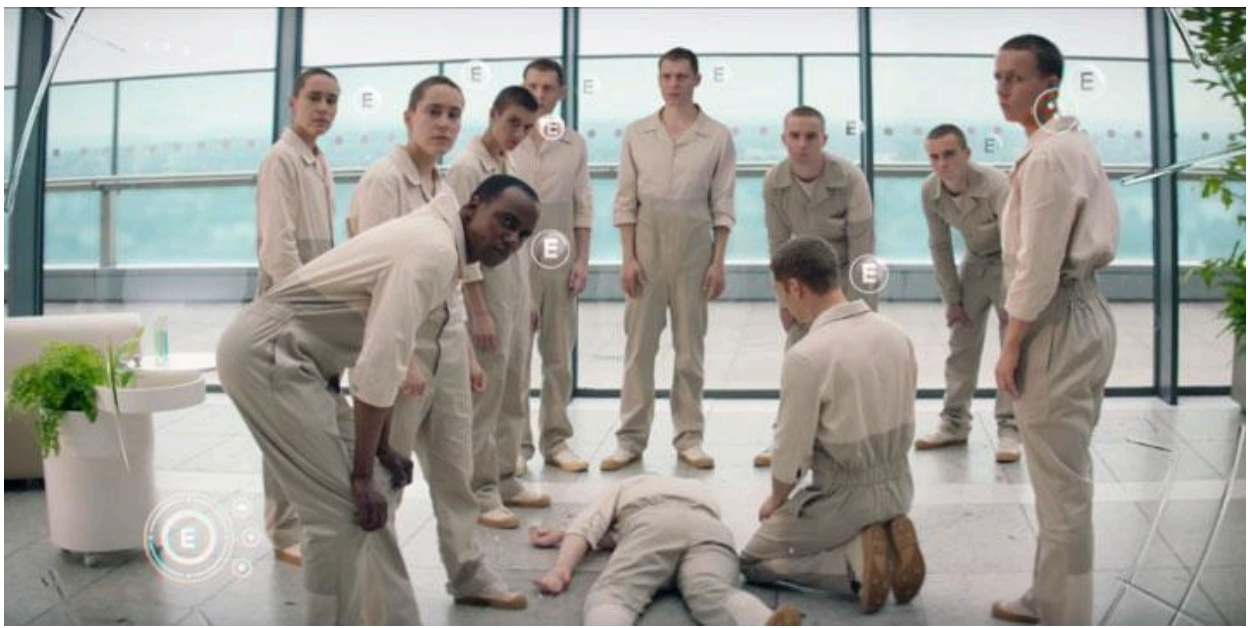

Figure 2, screenshot from Brave New World

By linking augmented reality to individuals' genetic status, this technology works analoglously to the AR encoding in *Brave New World*, as biometric population control based on eugenic principles. Historically, eugenics as an ideology and science, ideas about race, and systemic racism are inextricably intertwined and mutually constituent (Turda 2012). In both the TV series and the novel, society in Brave New World is governed by eugenic principles, where individuals are sorted into categories based on "high-value" and "low-value" genetic characteristics before they are even born. In Huxley's version, however, the genetic castes were also racialised, in that the higher-status people are encoded as white or pale-skinned, and the characters described as black are all ascribed to lower ranks, described using racist stereotypes (Rhines 2003). By contrast, the TV series portrays a seemingly post-racial society in which racialised physical features are not highlighted as meaningful, as is reflected in the casting. Instead of inferring human ability and value based on visible, racialised physical features, they go directly to the genetic source code. Although the hierarchy in the television version of *Brave New World* does not appear to be based on race, biometrics performs a similar truth-producing function in that it acts as a marker that points out the predetermined life cycle of bodies and defines their role as resources for society, rather than independent subjects. The hierarchy of genotypes, combined with the biometric labels displayed in the augmented reality thus has a racialising effect, even if society is portrayed as post-racial.

It is worth noting that the digital labels on the visual surface are not the only markers of genetic identity in *Brave New World*. Many of the inhabitants are clones of the same individual, and all the clones of the same type have the same biological appearance and are on the same level in the hierarchy. This means that their caste identity is already inscribed in their appearance. Linda Alcoff writes that "the reality of identities often comes from the fact that they are visibly marked on the body itself, guiding if not determining the way we perceive and judge others and are perceived and judged by them" (2006:21). Instead of being differentiated through the racialization of physical features, the bodies of the citizens in *Brave New World* are marked by virtue of resembling all the other clones of the same genotype – at least in terms of the lowest castes. In the higher status groups, there appears to be greater individual variety. For the lowest castes, the clones' identical appearance is a meaningful characteristic tied to the clones' collective rank, clearly indicating their position in the hierarchy at all times.

The bodies' norm-bound place in society is therefore overdetermined, both through appearance-based markers and the information conveyed via augmented reality. Alcoff (2006) points out that the visible is a sign and thus invites interpretation to discern what is behind what is immediately visible. The labels that neatly place everyone in categories replace this interpretation process with a template. The digital information layer is not necessary to determine an individual's position in the hierarchy; instead, it serves as a constant and indispensable reminder of the individuals' relevant characteristics and the framework of understanding into which they should be interpreted. A character's caste affiliation is doubly visible and is presented to the viewer fully interpreted. Here, too, augmented reality acts as a "grid of intelligibility", but in this context, it is used to indicate and maintain genetically conditioned hierarchical relationships in society. By revealing the truth about the body and its place in the hierarchy, the digital layer becomes a meaningful inscription of society's norms. The digital labels have a performative effect, in Butler's sense (2004), by acting as speech acts that produce and reproduce the norms they cite. Butler stresses that "we do not only act through the speech act; speech acts also act upon us" (2014:5), but in contrast to the usual understanding of speech acts as utterances or gestures performed by individuals, here the speech acts are inscribed in a digital visual layer that provides information about everyone. In Butler's description of the performativity of gender norms, change was possible by consciously or unconsciously misquoting the utterances that reproduce the gender norms, meaning the system could be undermined from within (Butler 2011). If we regard the augmented reality in *Brave New World* as a performative utterance, it is hard to imagine how these signifiers might be misquoted. By coding identities into a pervasive, centrally controlled digital layer that is integrated into the visual field, the hierarchical norm system is frozen.

**Power and manipulation**

The short film *Sight* (2012) presents a different version of augmented reality vision. The story follows a man called Patrick as he goes on a date with a young woman, Daphne. While they are eating, his field of vision is filled with details about her preferences and interests, evidently obtained from her social media, and helpful algorithmically generated suggestions on what he can do to win her over. His digitally enhanced charm offensive works, and she goes home with him. In his flat, she notices a trophy from a dating app digitally displayed on the wall, gets upset by the implications of having been played and gets up to leave. Patrick, who has helped develop the 'Sight' technology that integrates the digital vision, commands her to wait. We then see him hacking into her Sight interface while the woman remains standing, seemingly frozen in place.

In this example, it is not the state or an authoritarian power that is using technology to maintain control, and people do not come with predefined labels swirling around their heads like in *Anon* and *Brave New World.* The expressions of identity that can be read on the digital visual surface are largely user-defined, determined by how the individual chooses to present themself to the world and traces left by their digitally integrated lives. Instead of power being exercised through a pre-defined category system where everyone is placed in given identities, in this example, the power lies in the algorithm, which tells Patrick how to manipulate Daphne.

"We see the world differently when we view it with algorithms," writes Jill Rettberg (2019), but the algorithmic gaze that fills Patrick's field of vision will be familiar to many. In 1989, Laura Mulvey (1989) introduced the concept of the "male gaze", to characterize how in Hollywood films women are represented as sexual objects for the male spectator's pleasure. This perspective is not limited to cinema; it is also frequently found in other visual media, including how we represent ourselves and others in social media (Oliver 2017). The algorithm in this short film retrieves information from Daphne's social media, as well as her physical presence, and recontextualises it – everything she does and is is transformed into information about her sexual availability. Instead of being integrated into the camera perspective, the male gaze is encoded into the algorithm that reads Daphne's facial expressions and interprets her preferences from her digital profiles, telling Patrick what he needs to do to get her into bed. The augmented reality game Patrick is playing while he is on a date with Daphne stages him as the active party in a heterosexual seduction game in which she is the prize. Daphne is reduced to a challenge to be decoded and won.

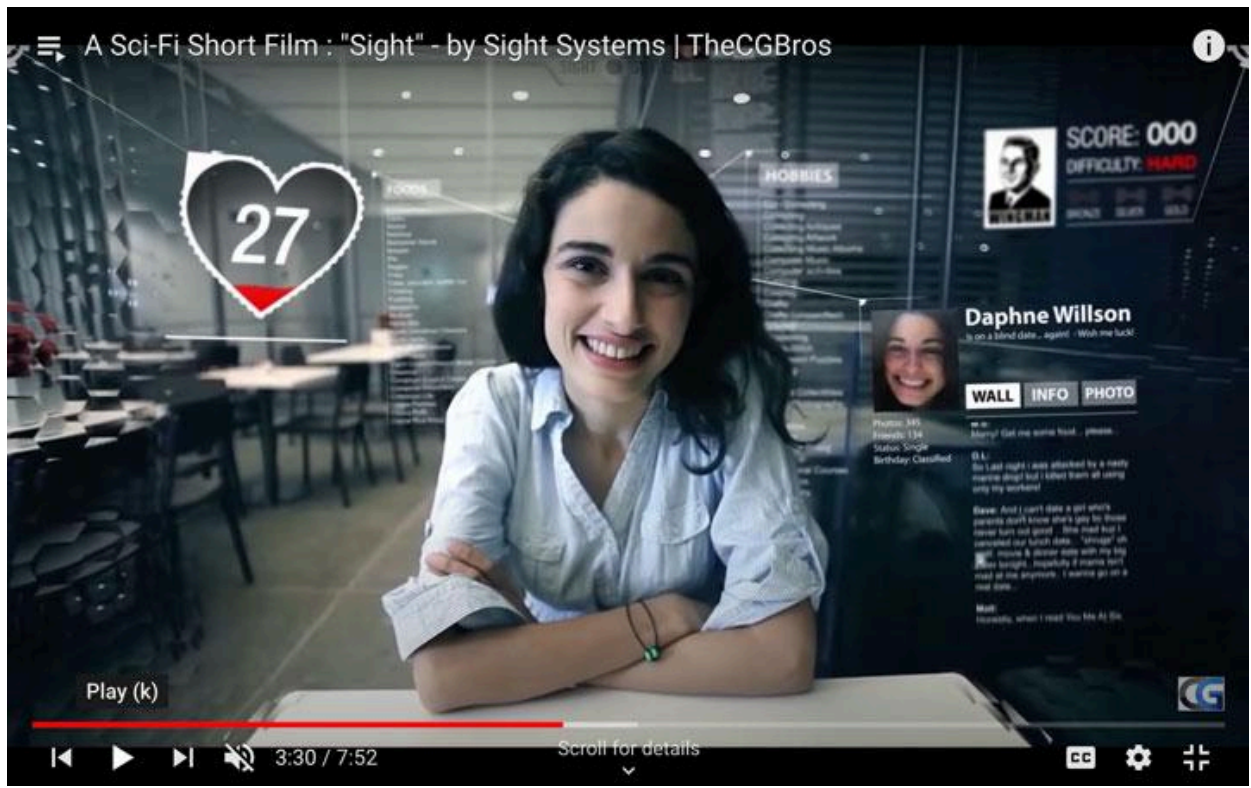

Figure 3, screenshot from Sight

Although this may seem unrealistic, the technology shown in this video, in which artificial intelligence continuously offers advice on ongoing social interactions, is one of the more realistic representations of augmented reality today. Many of the hits in a literature search about current uses of augmented reality discuss the use of AR to help autistic children learn "normal" social interaction codes and behaviour, and programs using smartglasses to offer this type of assistance are already in use (Chen et al. 2015; Liu et al. 2017; Lorenzo et al. 2019; Sahin et al. 2018). One of them, the *"Empowered Brain"* system, is used in conjunction with Google Glass glasses, encouraging the user to meet the gaze of the person they are talking to and correctly guess what emotions their facial expression reveals (Sahin et al. 2018). In other words, these are smartglasses that, in conjunction with artificial intelligence, encourage the user to behave in line with normative social behaviour – behaviour that will be perceived as comfortable and normal for the other person. The dating assistant app that is used in *Sight* does much of the same: encouraging the user to smile at the right time, reading the facial expressions of the woman he is conversing with and letting him know when they have touched upon a sore point that makes her uncomfortable, but it also lets him know when she has drunk the "optimal amount of alcohol" to be receptive to a proposal to go home with him. Although these are two completely different uses of augmented reality, they are both expressions of what Lee and Larsen (2019) call algorithmic normativity.

As early as 1995, Allison Adam argued in her feminist critique of artificial intelligence that the knowledge encoded into algorithmic systems would reflect existing hierarchies in which women's perspectives and participation were in danger of being excluded. She pointed out that "AI can be used to exclude the other, the different and inevitably women" (Adam 1995:19), which has proved to be a well-founded warning. Over the past decade, there have been numerous examples of algorithms that (re)produce gender discrimination and/or stereotypical representations of marginalised identities (Benjamin 2019; Bolawumni and Gebru 2018; Garcia 2016). In *Sight,* the algorithmic gaze excludes Daphne's perspective and agency, shaping a dynamic in which her subjectivity is set aside in favour of a portrayal of her as a sexual object.

Taina Bucher (2018) argues that algorithms are part of social power structures through what she calls "programmed sociality", meaning that algorithms organise and shape social formations, and that the algorithms, in turn, are expressions of social norms and power structures (Bucher 2018:15). Building on Sara Ahmed's approach to orientations as something that shapes how the world takes shape around us, she points out that algorithms orient us by creating an outside world where some subject positions are more possible and accessible than others. "The algorithm gets suffused with agendas and assumptions", Bucher writes (2018:161), as is illustrated in an almost caricature-like fashion in the short film *Sight*. The algorithm in the dating game that unfolds in Patrick's field of vision places his interactions in a context structured by the orientation inscribed into this algorithm, in which women are encoded as objects of seduction – tempting puzzles that men achieve points for solving. This dynamic is played out in his encounter with Daphne, who is unaware of the game in which she has become a passive character. These archetypal subject positions, the man as the active seducer and the woman as the unwitting victim, and the associated power relations, in which he is the active agent and she is being subjected to his actions, are coded into the augmented reality that *Sight* portrays. The algorithm in the dating game structures the interactions between Patrick and Daphne in line with the perspective of the male gaze, and by extension, it enables and normalises men's abuse of women.

Again, this can be understood as performative utterances, which reiterate and reproduce existing gender norms in their staging of the interpersonal encounter. Building on the assumption that norms are maintained by performative repetition of speech acts, this is an example where algorithmic augmented reality plays an active part in playing out these utterances in ongoing social interaction. In this context, augmented reality serves as a normative and performative script, a grid of intelligibility that appears not only in Patrick's field of vision, but also in the viewer's (Butler 2004). On cinematography, Mulvey wrote that "going far beyond highlighting a woman's to-be-looked-at-ness, cinema builds the way she is to be looked at into the spectacle itself" (1989:17). When this principle is coded into augmented reality, everyday interactions are also structured according to the same principle. This is at the heart of how power unfolds through the augmented reality in all three works: by framing and contextualising the objects in the field of vision, it shapes the gaze that sees.

**Augmented reality, decreased agency?**

Building on the readings of these three works through the lens of feminist theory, some patterns start to emerge. Each of these works addresses tendencies that we can already glimpse in existing digital technologies and current online culture, and they imagine how these traits might develop in an augmented reality organised according to the same principles. The first of these tendencies is the use of labels that stabilise and publicise identities. The augmented reality in *Anon* acts as a technological filter, where everyone is identified through reductive categories or digital labels. Interpreting these labels as performative speech acts, it becomes clear how augmented reality helps to set the framework for possible and impossible subject positions. Through forced visibility, these labels act as an intelligibility filter – a system through which identities are stabilised and defined. At the same time, this system draws a line between the people who are safely placed in society and the unintelligible and unrecognisable "others", laying the foundation for increasingly fine-meshed surveillance.

The second tendency that I want to highlight is the use of these labels or categorisation systems to shape and maintain societal hierarchies by defining how bodies can be read. In line with the ideology of eugenics, augmented reality in *Brave New World* is used to mark internal boundaries in society and support biologically rooted hierarchical stratification. This example unveils the risk that augmented reality might be used to stage bodies in the field of vision based on a racist or biologically deterministic logic in which digital information takes precedence over other possible readings of bodies.

The third tendency that emerged in my readings is algorithmic normativity's role in augmented reality designed to aid in social situations. In *Sight*, augmented reality is combined with artificial intelligence to interpret an ongoing social situation, giving one party an upper hand. The digital information and game logic on which the program is based, structures the social situation in line with the male gaze and establishes an algorithmic normativity in which the subject positions in interpersonal relationships are defined in advance. There is therefore reason to examine which norms are encoded into the algorithmic gaze when augmented reality is invited to regulate or influence social interactions.

What these popular cultural imaginings of augmented reality envisage is that the logic that is encoded into the systems included in our digital everyday life today will also be encoded into the AR field of vision, where they will be able to frame and pre-define how we relate to the objects and people we look at, and at the same time expose us to the gaze of those in power. *Anon*, *Brave New World* and *Sight* all present a world in which augmented reality helps shape how people sense the outside world, and where the categories and frameworks of understanding that are activated when one turns one's gaze to another person have been predetermined. In all three works, normative categories are imported into the field of vision in a way that funnels what is sensed into established power structures.

In this sense, augmented reality works like what N. K. Hayles (2017) dubs "non-conscious cognition" to describe all the automatic but meaningful processes that occur outside our consciousness, such as processing sensory impulses to form meaningful impressions, orientation in time and space, reflexive reactions, automatic inferences, and muscle memory. Hayles points out that non-conscious cognition is a capacity humans share with other life forms and technological devices, such as algorithms. Humans can thus be part of an "assemblage" with other cognitive actors, where cognition is carried out in a distributed network through which meaning is exchanged among humans, machines and other organisms (Hayles 2017). However, this does not mean that people are part of a network with symmetrical power relations or that all the actors have the same degree of agency. From this perspective, the augmented realities depicted in these works can be regarded as actors that impart meaning asparts cognitive assemblages with the people sensing the world through it. The result is an automation of shared knowledge and preconceptions (Andrejevic 2015). The information that is incorporated into the field of vision serves as what Mark Andrejevic describes as a "prosthetic collective unconscious" (2015:xvi), while the cognitive process that produces it is located in an assemblage of people, discourses and technology, where none of the actors necessarily have a complete overview. Within these assemblages, augmented reality takes on a performative and norm-bearing role by forming a grid of intelligibility that scripts or structures social interactions (Butler 2014).

All three analyses demonstrate that augmented reality cannot be discussed as an isolated phenomenon independent of other digital and societal structures. Augmented reality will necessarily be combined with the formations already shaping our experience of the digital sphere, such as surveillance structures, classification systems and algorithmic processes. It is therefore not possible to understand augmented reality without taking into account how it is included in assemblages with other technologies, systems of meaning and institutions.

Although these works deal with augmented reality in different ways, in different contexts, and with their distinct thematic focus, they all show how augmented reality can be included in assemblages with other actors where they serve to uphold power relations, reproduce norms, stabilise identities and establish subject positions. John Novak writes that the goal of augmented reality is "to increase human agency and a human's sense of agency" (2016:27), but perceived agency and actual agency are not the same. There is reason to ask whose agency is being expanded, who will have their agency curtailed by a given implementation of augmented reality, and how this is impacted by and problematised in popular culture and science fiction.

Finally, it is worth noting that the arguments presented in this analysis are derived from existing tendencies and features of social media and popular cultural visions. Science fiction can be read as imagining the future, but also as diagnosing the present. In the latter perspective, the visions of augmented reality analysed in this text may also serve as a lens to help illuminate the present.